\documentclass[conference]{IEEEtran}
\usepackage{setspace}
\usepackage{clrscode}
\usepackage{amsmath}
\usepackage{amssymb}
\usepackage[dvips]{graphicx}
\usepackage{epsfig}
\usepackage{hyperref}
\usepackage{bm}
\usepackage{subfigure}
\usepackage{indentfirst}
\usepackage{float}
\usepackage{bm}

\newcommand{\figsize}{0.38}
\newcommand{\C}{\mathbb{C}}

\newcommand{\E}{\mathbb{E}}

\newcommand{\tabincell}[2]{\begin{tabular}{@{}#1@{}}#2\end{tabular}}

\newtheorem{Lem1}{Proposition}
\newtheorem{Lem}{Theorem}
\newtheorem{Rem}{Remark}

\newtheorem{Corr}{Corollary}

\begin{document}
\title{Performance Analysis of Indoor THz Communications with One-Bit Precoding}
       \author{
\IEEEauthorblockN{Dan Li\IEEEauthorrefmark{1}, Deli Qiao\IEEEauthorrefmark{1}, Lei Zhang\IEEEauthorrefmark{2}, and Geoffrey Ye Li\IEEEauthorrefmark{3} }
\IEEEauthorblockA{\IEEEauthorrefmark{1}\small{School of Information Science and Technology, East China Normal University, Shanghai, China}}
\IEEEauthorblockA{\IEEEauthorrefmark{2}\small{School of Engineering, University of Glasgow, Glasgow, U.K.}}
\IEEEauthorblockA{\IEEEauthorrefmark{3}\small{School of Electrical and Computer Engineering, Georgia Institute of Technology, Atlanta, Georgia}}
\small{Email: 51161214012@stu.ecnu.edu.cn, dlqiao@ce.ecnu.edu.cn, lei.zhang@glasgow.ac.uk, liye@ece.gatech.edu}
}

\maketitle

\begin{abstract}
 In this paper, the performance of indoor Terahertz (THz) communication systems with one-bit digital-to-analog converters (DACs) is investigated. Array-of-subarrays architecture is assumed for the antennas at the access points, where each RF chain uniquely activates a disjoint subset of antennas, each of which is connected to an exclusive phase shifter. Hybrid precoding, including maximum ratio transmission (MRT) and zero-forcing (ZF) precoding, is considered. The best beamsteering direction for the phase shifter in the large subarray antenna regime is first proved to be the direction of the line-of-sight (LOS) path. Subsequently, the closed-form expression of the lower-bound of the achievable rate in the large subarray antenna regime is derived, which is the same for both MRT and ZF and is independent of the transmit power. Numerical results validating the analysis are provided as well.
 \end{abstract}

 \section{INTRODUCTION}

Due to the exponential increase in mobile traffic, advanced transmission techniques have been developed to support the demand in high data rate in next generation communication systems \cite{5gtechnique}-\cite{lulu-mmimo}. It is expected that more spectral bands will be required to support future wireless communications. Amongst other, the Terahertz (THz) band (0.1 - 10 THz) is proposed as one of the promising solutions to enable ultra-high-speed communications \cite{thz}-\cite{thzrate}. The available bandwidth in the THz band is more than one order of magnitude over the microwave frequency. Such large and unoccupied frequency resources can be utilized to address the spectrum scarcity and capacity limitation of current communication systems. To address the high path loss and molecular absorption, a large number of antennas are required for THz communications \cite{thzrate}, which is also easy to realize in a very small area due to the very short wavelength of THz signals.

 Modeling and analysis of THz communications have attracted increasing interests recently (see, e.g., \cite{ino}-\cite{clincomm} and references therein). For instance, a new propagation model incorporating the molecular absorption in the Terahertz band has been developed in \cite{ino}. The indoor Terahertz channels in range of 275 GHz to 325 GHz has been characterized in \cite{inp}. Distance-aware multi-carrier transmission by exploiting the distance frequency dependent transmission windows in THz band has been proposed in \cite{daba}. The indoor THz communications with antenna subarrays for single users with perfect channel state information have been analyzed in \cite{indoor}, while the performance for multi-users with partial channel state information has been characterized in \cite{ine}. Based on the analysis \cite{clincomm}, when hybrid precoding and an array-of-subarrays architecture are employed, both the spectral and energy efficiency in the THz communications can be improved. In the above analysis, high-resolution digital-to-analog converters (DACs), which may be high-cost and power-inefficient, are assumed in the system and therefore quantization error is ignored.

We note that quantized precoding for massive MIMO in the micro-/millimeter-wave communication systems with low-resolution DACs has received much interest recently (see, e.g., \cite{quantmimo}-\cite{eeonebit}, and references therein). For instance, it has been shown in \cite{quantmimo} showed that performance of a massive MIMO system with low-resolution DACs of 3-4 bits is close to that with infinite-resolution DACs, where each antenna is driven by one radio frequency (RF) chain. In \cite{eeonebit}, the energy efficiency for mmWave massive MIMO precoding with low-resolution DACs has been analyzed. Hybrid precoding schemes with both fully-connected structure, in which each RF chain drives all antennas with individual phase-shifters, and partially-connected structure, in which each RF chain is linked to only a subarray of antennas with individual phase shifters, have been considered using AQN model \cite{aqn} as the quantization error model, which is suitable in the low-power regime. It has been shown in \cite{eeonebit} that partially connected structure can achieve better energy efficiency and increasing the resolution of DACs doesnot significantly improve spectral efficiency.

In this paper, we investigate the indoor THz communications with one-bit DACs. We incorporate the distance aware multi-carrier transmission strategy to support multi-user transmissions in the THz band. We adopt the array-of-subarrays structure, where each RF chain drives only a subarray of antennas with individual phase shifters. We use the hybrid precoding scheme in which users are first divided into groups based on beam steering angles confined by the THz channel characteristics in the analog domain. We assume that users in the same group are assigned orthogonal frequencies based on the distance-aware multi-carrier scheme with digital precoding in the baseband. We consider the large subarray antenna regime and show that the analog beamforming angles for each group are decided by the line-of-sight (LOS) path. We derive the closed-form expression for the lowerbound on the achievable rate. We also investigates the impact of phase uncertainties on the rate degradation in the single-user case.

This paper is organized as follows. In Section II, the system model and preliminaries on indoor THz channel model and hybrid precoding are discussed. Main results on the analysis in the large subarray antenna regime with one-bit DACs are presented in Section III, with numerical results given in Section IV. Finally, Section V concludes this paper with some lengthy proofs in Appendices.

\emph{Notation:} $\mathbf{C}^{T}$, $\mathbf{C}^{H}$, and $tr(\mathbf{C})$  denote the transpose, Hermitian and trace of matrix $\mathbf{C}$. $\text{diag}(\mathbf{C})$ denotes the main diagonal of the matrix $\mathbf{C}$, while $\text{diag}(\mathbf{c})$ represents the diagonal matrix generated by the vector $\mathbf{c}$.  $[\mathbf{C}]_{\iota,\ell}$ represents the entry on the $\iota$-th row and the $\ell$-th column. $\mathbb{E}[\cdot]$ is the expectation and $\delta (\cdot)$ is the Dirac delta function.

 \section{System Model and Preliminaries}\label{sec:model}
In this section, we will first briefly introduce the system model, and then the indoor THz channel model and hybrid precoding scheme with one-bit DACs, which are useful for the subsequent analysis.

\begin{figure}
    \centering
    \includegraphics[width=\figsize\textwidth]{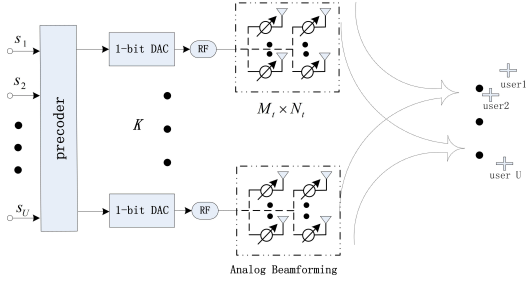}
    \caption{System model.}
    \label{fig:system model}
\end{figure}

\subsection{System Model}

We consider the THz communication systems with one-bit DACs as illustrated in Fig.1. We assume that the array-of-subarrays architecture is the same as in \cite{clincomm}, which is one kind of hybrid precoding structure and has been shown to improve both the spectral and energy efficiency in the Terahertz band with reduced complexity. In particular, the Access Point (AP) is equipped with $K$ antenna subarrays, each of which is composed of $M_{t} \times  N_{t}$ tightly-packed antenna elements. Each antenna element is driven by an analog phase shifter. At baseband, each subarray is connected to one RF chain and an one-bit DAC. The AP is assumed to serve $U$ users with different distances, each is equipped with only one subarray of size $M_{r} \times  N_{r}$. In general, $K$ is assumed to be greater than $U$ to offer sufficient degrees of freedom to all users.

Denote $a$ as the subarray antenna element spacing. Usually, $a$ is assumed to be less than the wavelength while the spacing between the adjacent subarrays is much larger than the wavelength. In this way, independent channels can be expected for different subarrays. Additionally, the distance from any user to the AP is assumed to be much larger than the spacing between the subarrays such that the distance between user $i$ and any subarray is the same and denoted as $d_i$.

 \subsection{Indoor Terahertz Channel Model}

The multi-ray channel response for one antenna subarray is given by \cite{indoor}
 \begin{align}\label{eq:channel}
 &\mathbf{F}_{sub}(f,d)\nonumber\\
 =&\sqrt{M_{t}N_{t}M_{r}N_{r}}\big[\eta_{L}(f,d)\Omega_{t}\Omega_{r}\mathbf{a}_{r}(\psi_{L}^{r},\phi_{L}^{r})\mathbf{a}_{t}^{H}(\psi_{L}^{t},\phi_{L}^{t})\nonumber\\
 +&\sum_{i=1}^{J}\eta_{i}(f,d)\Omega_{t}\Omega_{r}\mathbf{a}_{r}(\psi_{i}^{r},\phi_{i}^{r})\mathbf{a}_{t}^{H}(\psi_{i}^{t},\phi_{i}^{t})\big],
\end{align}
where $J$ is the number of non-LOS (NLOS) rays, $\eta(f,d)=|\eta(f,d)|e^{j\vartheta}$ denotes the channel coefficient of the ray with $\vartheta$ representing the independent phase shift of each path, and $\psi^{t},\phi^{t}$ and $\psi^{r},\phi^{r}$ are the corresponding azimuth and elevation angles of departure and arrival (AoD/AoA) for the rays, respectively. $\Omega_{t}$ and $\Omega_{r}$ are the transmit and receive antenna gains. $\mathbf{a}_{r}(\psi^{r},\phi^{r})$ and $\mathbf{a}_{t}(\psi^{t},\phi^{t})$ are the subarray steering vectors at the transmit and receive sides, respectively. For an $(M,N)$-element uniform planar array, the array steering vector is given by
\begin{align}
  \mathbf{a}(\psi,\phi)=\frac{1}{\sqrt{MN}}[1,\cdots, e^{j\frac{2\pi a}{\lambda}[m\cos\psi\sin\phi+n\sin\psi\sin\phi]},\nonumber\\
\cdots, e^{j\frac{2\pi a}{\lambda}[(M-1)\cos\psi\sin\phi+(N-1)\sin\psi\sin\phi]}]^{T},
\end{align}
where $m$ and $n$ are the antenna element indexes with $0\leq m\leq M-1, 0\leq n\leq N-1 $, $\lambda$ is the wavelength and $a$ is the antenna element spacing.

The THz signals are sensitive to the atmospheric attenuation and molecular absorption. The path gain of the line-of-sight (LOS) path is expressed as \cite{ino}
\begin{align}
 |\eta_{L}(f,d)|^{2}&=\mathfrak{L}_{asp}(f,d)=\left(\frac{c}{4\pi fd}\right)^{2}{e}^{-\kappa_{abs}(f)d}.
\end{align}
where $c$ is the speed of light in free space, $f$ is the carrier frequency, $d$ is the path length, and $\kappa_{abs}$ is the frequency-dependent medium absorption coefficient, which is decided by the transmission medium at a molecular level. On the other hand, due to the sub-millimeter wavelength of the THz signals, the indoor surfaces are rough at THz frequency, which will introduce additional loss of the NLOS rays \cite{inp}. Then, the path gain of the $i$-th NLOS ray with one reflection satisfies
\begin{align}\label{eq:nlosgain}
 |\eta_{i}(f,d)|^{2}=\Gamma_{i}^{2}(f)\mathfrak{L}_{asp}(f,d),
\end{align}
where $\Gamma$ is the reflection coefficient composed of the Fresnel reflection coefficient and the Rayleigh roughness factor. Note that only up to two reflections are considered in the THz band due to the high reflection losses \cite{inp}, and the associated path gain can be written similarly to (\ref{eq:nlosgain}). Therefore, the THz channel has limited number of paths.

In the THz communication system, the propagation will be affected by the molecular absorption severely in certain frequencies due to the molecular absorption. Then, different transmission windows, each with different bandwidth varying with the communication distance, are created and incorporated as an important feature in designing the transmission strategies in the THz band \cite{daba}. For instance,  distance-aware multi-carrier transmission has been proposed and analyzed for indoor THz communications, where the transmission windows, 0.6-0.7 THz  and 0.8-0.95 THz, will be assigned  to the user with 10 m distance to the AP, while the transmission windows, 0.5-0.6 THz, 0.7-0.8 THz, and 0.95-1 THz, will be allocated to the user 1 m away. In this paper, we adopt such distance-aware multi-carrier transmission, and denote the subcarrier index as $w$ with central frequency $f^w$ and bandwidth $B$ and $\mathcal{W}_u$ as the set of subcarrier index for user $u$.

\subsection{Hybrid Beamforming and Transmission Rate}

We consider hybrid beamforming strategy, in which digital beamforming at baseband and analog beamforming at each antenna subarray are performed. In this scheme, the users are first divided into user groups based on the transmit analog beamforming angles, and then after obtaining the effective channel information of the subarrays with the analog beamforming, digital precoding such as maximal ratio transmission (MRT) and zero-forcing (ZF) are applied.

\subsubsection{Analog Beamforming}
Specifically, subarrays activate a set of antennas to produce a beam pre-scanning with an angular separation $\varphi$, the different users in the same angle section are viewed as one group. Denote the user set in group $q$ as $\mathcal{U}_{q}$. The users in the same group share the set of antenna subarrays $\mathcal{K}_q$. Let $(\psi_{0}^{t}, \phi_{0}^{t})$ be the transmit analog beamforming angle of the subarray $k\in\mathcal{K}_q$. Suppose that $(\psi_{0}^{r}, \phi_{0}^{r})$ is the receive analog beamforming angle of user $u\in \mathcal{U}_{q}$. Then, the effective channel between the user and the $k$-th subarray of the AP is given by
\begin{align}\label{eq:equivchannel1}
h_{k}(f_{u}^{w},d_{u})&=\mathbf{a}_{r}^{H}(\psi_{0}^{r},\phi_{0}^{r})\mathbf{F}_{sub_k}(f_{u}^{w},d_{u})\mathbf{a}_{t}(\psi_{0}^{t},\phi_{0}^{t}),
\end{align}
where $d_{u}$ is the distance from the AP to user $u$.

The transmit beamforming angles for the users in group $q$, $(\tilde{\psi}_{q}^{t},\tilde{\phi}_{q}^{t})$, and receive beamforming angles for user $u$ in group $q$, $(\tilde{\psi}_{u,q}^{r},\tilde{\phi}_{u,q}^{r})$, can be selected from the codebook as follows
\begin{align}\label{eq:equihbg}
&\{(\tilde{\psi}_{q}^{t},\tilde{\phi}_{q}^{t}),(\tilde{\psi}_{u,q}^{r},\tilde{\phi}_{u,q}^{r})\}\nonumber\\
&={\arg\max}_{\substack{(\psi_{q}^{t},\phi_{k,q}^{t})\in\Theta^t, \\(\psi_{u,q}^{r},\phi_{u,q}^{r})\in\Theta^r}}\sum_{u\in \mathcal{U}_{q}}\sum_{w\in \mathcal{W}_{u}}\frac{\parallel \mathbf{h}(f_{u}^{w},d_{u})\parallel^{2}}{|\eta_{L}(f_{u}^{w},d_{u})|^{2}},
\end{align}
where $\mathbf{h}(f_{u}^{w},d_{u})$ is the effect channel between the AP and user $u$ and with element $h_{k}(f_{u}^{w},d_{u})$ for $k\in \mathcal{K}_q$, $\Theta^t$ and $\Theta^r$ denote the transmit and receive beamforming codebook, respectively. Denote $\tilde{\mathbf{h}}(f_{u}^{w},d_{u})$ as the effective channel of one antenna subarray for user $u$ in group $q$ with analog beamforming angles obtained from (\ref{eq:equihbg}).

According to the distance-aware multi-carrier transmission strategy, there might be different numbers of users in subcarrier $w$. Denote $\mathcal{U}^w$ as the set of users sharing the same carrier with central frequency $f_u^w$ with total number $U^w$. Let $q_u$ be the group index associated with the users $u\in\mathcal{U}^w$. Then, we know that $\mathcal{K}^w=\cup_{u\in\mathcal{U}^w}\mathcal{K}_{q_u}$ represents the set of subarrays transmitting signals in subcarrier $w$ with total number $K^w=\sum_{u\in\mathcal{U}^w}K_{q_u}$.

Let $\mathbf{F}_{u}=[\mathbf{F}_{{1},u},\mathbf{F}_{{2},u},\ldots,\mathbf{F}_{K^{w},u}]\in\mathbb{C}^{M_rN_r\times K^wM_tN_t}$ be the channel matrix for user $u\in\mathcal{U}^w$ with respect to the subarrays in $\mathcal{K}^w$. Then, the effective channel vector $\tilde{\mathbf{h}}_{u}(f_{u}^{w},d_{u})\in\mathbb{C}^{1\times K^w}$ for user $u$ in group $q$ with analog beamforming is given by
\begin{align}\label{eq:equivah}
\tilde{\mathbf{h}}_{u}(f_{u}^{w},d_{u})=\mathbf{V}_{u}^{H}\mathbf{F}_{u}\mathbf{D},
\end{align}
where $\mathbf{V}_u=\mathbf{a}_{r}(\tilde{\psi}_{u,q}^{r},\tilde{\phi}_{u,q}^{r})\in\mathbb{C}^{M_rN_r\times1}$ represents the receive analog beamforming vector, and $\mathbf{D}\in\mathbb{C}^{K^w M_tN_t\times K^w}$ stands for the transmit analog beamforming operation and is a block matrix with diagonal components given by $[\mathbf{a}_{t}(\tilde{\psi}_{1}^{t},\tilde{\phi}_{1}^{t}),\ldots,\mathbf{a}_{t}(\tilde{\psi}_{K^w}^{t},\tilde{\phi}_{K^w}^{t})]$ corresponding to the transmit beam steering vectors for subarrays $k\in\mathcal{K}^w$ and other components all being zero matrix of size $M_tN_t\times1$. We assume that $\tilde{\mathbf{h}}_{u}(f_{u}^{w},d_{u})$ can be perfectly estimated at the user side and fed back to the AP. The equivalent channel matrix for all users can be expressed as
\begin{align}\label{eq:hmat}
\tilde{\mathbf{H}}&=[\tilde{\mathbf{h}}_{1}^T,\ldots,\tilde{\mathbf{h}}^T_{U^w}]^{T}.
\end{align}

%

\subsubsection{Quantized Precoding}

At the baseband, we consider quantized precoding by incorporating the one-bit DACs at the transmitter side. We model the precoding operation as \cite{quantmimo}\cite{inh}
\begin{align}
\mathbf{x} = {\L}(\mathbf{Q}\mathbf{s})+\mathbf{d}=\mathbf{G}\mathbf{Q}\mathbf{s} + \mathbf{d},
\end{align}
where $\mathbf{x}\in\mathbb{C}^{U^w\times 1}$ is the precoded transmit signal, $\mathbf{s}\in\mathbb{C}^{U^w\times1}$ is the Gaussian source data with $\E\{\mathbf{s}\mathbf{s}^{H}\}=\mathbf{I}_{U^w}$, ${\L}(\cdot)$ represents the quantizer-mapping function, $\mathbf{Q}\in \C^{K^w\times U^w}$ denotes the linear precoding operation performed at the baseband with $tr\{\mathbf{Q}\mathbf{Q}^H\}=P^w$ where $P^w$ is the transmit power for subcarrier $w$, $\mathbf{d}\in\mathbb{C}^{K^w\times1}$ stands for the distortion introduced by the DACs, and $\mathbf{G}\in \C^{K^w\times K^w}$ is the diagonal matrix modeling the operation of the one-bit DACs quantization, which is given by
 \begin{align}\label{eq:quantmat}
 \mathbf{G}=\sqrt{\frac{2P^w}{\pi K^w}} \text{diag}(\mathbf{Q}\mathbf{Q}^{H})^{-\frac{1}{2}}.
 \end{align}
When MRT is adopted at baseband, we have
\begin{align}\label{eq:mrt}
  \mathbf{Q}_{MRT}=\sqrt{\frac{P^{w}}{tr(\tilde{\mathbf{H}}\tilde{\mathbf{H}}^{\mathbf{H}})}} \tilde{\mathbf{H}}^{\mathbf{H}}.
  \end{align}
When ZF is adoted at baseband, we have
\begin{align}\label{eq:zf}
 \mathbf{Q}_{ZF}=\sqrt{\frac{P^{w}}{tr((\tilde{\mathbf{H}}\tilde{\mathbf{H}}^{H})^{-1})}} \tilde{\mathbf{H}}^{H} (\tilde{\mathbf{H}}\tilde{\mathbf{H}}^{H})^{-1}.
 \end{align}

Let $\mathbf{q}_{i}$ be the $i$-th column of the precoding matrix $\mathbf{Q}$. Then, the received signal $y_u$ after receive analog beamforming  at user $u$ in group $q$ in subcarrier $w$ can be described as
\begin{align}
y_u &= \tilde{\mathbf{h}}_{u}(f_{u}^{w},d_{u})\mathbf{G}\mathbf{q}_u s_u +\sum_{u'\in\mathcal{U}^w,u'\neq u}\tilde{\mathbf{h}}_{u}(f_{u}^{w},d_{u})\mathbf{G}\mathbf{q}_{u'}s_{u'} \nonumber\\
&+ \tilde{\mathbf{h}}_{u}(f_{u}^{w},d_{u})\mathbf{d} + n_u,
\end{align}
where $\mathbf{q}_u\in\mathbb{C}^{K^w\times1}$ denotes the digital precoding vector corresponding to user $u$, and $n_u$ is the additive Gaussian noise with power $N_0$.

Then, we know that the achievable rate of user $u$ with one-bit DACs is lowerbound by \cite{quantmimo}
\begin{small}
\begin{align}\label{eq:rate}
 R_{u}\ge &\mathbb{E}\Big[\sum_{w\in{\mathcal{W}_u}}B\log_{2}(1+ \nonumber\\
 &\hspace{-1cm}\frac{|\tilde{\mathbf{h}}_{u}(f_{u}^{w},d_{u})\mathbf{G}\mathbf{q}_{u}|^{2}}{\sum_{u\neq{i}}|\tilde{\mathbf{h}}_{u}(f_{u}^{w},d_{u})\mathbf{G}\mathbf{q}_{i}|^{2}+\tilde{\mathbf{h}}_{u}(f_{u}^{w},d_{u})\mathbf{C}_{dd}\tilde{\mathbf{h}}^H_{u}(f_{u}^{w},d_{u})+N_{0}})\Big],
\end{align}
\end{small}where $\mathbf{C}_{dd}=\mathbb{E}[\mathbf{d}\mathbf{d}^{H}]$ stands for the covariance of the distortion $\mathbf{d}$. For one-bit DACs, the closed form of $ \mathbf{C}_{dd}$ is given by \cite{inh}, \cite{dacerror}
\begin{align}\label{eq:distcov}
\hspace{-.5cm}\mathbf{C}_{dd} &=\frac{2P^{w}}{\pi K^w}\{\arcsin(\text{diag}(\mathbf{Q}\mathbf{Q}^{H})^{-\frac{1}{2}} \Re\{\mathbf{Q}\mathbf{Q}^{H}\} \text{diag}(\mathbf{Q}\mathbf{Q}^{H})^{-\frac{1}{2}}) \nonumber\\
 &+j \arcsin(\text{diag}(\mathbf{Q}\mathbf{Q}^{H})^{-\frac{1}{2}}\Im\{\mathbf{Q}\mathbf{Q}^{H}\}\text{diag}(\mathbf{Q}\mathbf{Q}^{H})^{-\frac{1}{2}})\}\nonumber\\
 &-\mathbf{G}\mathbf{Q} \mathbf{Q}^{H} \mathbf{G}^{H}.
\end{align}

The sum rate of the indoor THz communication systems is then given by
\begin{align}
R=\sum_{u=1}^{U} R_{u}.
\end{align}

\section{Performance Analysis}\label{sec:analysis}

In this section, we investigate the achievable rate of the indoor THz communications with one-bit DACs in the large subarray antenna regime, in which the number of antennas for each subarray goes to infinity.

Due to the high reflection loss, the power of the first-order reflected path is attenuated by more than $10$ dB on average and the second-order reflection by more than $20$ dB compared with the path gain of LOS path in the THz propagation \cite{inp}, i.e., $|\eta_L|>|\eta_i|,\,\forall i$. High directionality in the THz band can be expected, we first have the following result.
\begin{Lem1}\label{prop:angles}
When the number of antennas for each subarray goes to infinity, the transmit and receive beamforming angles for user $u$ in group $q$ are given by
\begin{align}\label{eq:angles}
\{(\tilde{\psi_q^{t}},\tilde{\phi}_q^{t}),(\tilde{\psi}_{u,q}^{r},\tilde{\phi}_{u,q}^{r})\}=\{(\psi_{u,L}^{t},\phi_{u,L}^{t}),(\psi_{u,L}^{r},\phi_{u,L}^{r})\}.
\end{align}
\end{Lem1}
\emph{Proof:} See Appendix \ref{app:angles} for details. $\hfill\square$
\begin{Rem}
From Proposition \ref{prop:angles}, one simple and effective way to determine the beamforming angles in the analog beamforming domain is to select the AoD/AoA of the LOS path when the number of antennas in each subarray goes to infinity.
\end{Rem}

Henceforth, we assume that the analog beamforming angles are determined through Proposition \ref{prop:angles}. As a result, when the number of subarray antennas goes to infinity, (\ref{eq:equivah}) can be rewritten as
\begin{align}\label{eq:infchannel}
\tilde{\mathbf{h}}_{u}(f_{u}^{w},d_{u})\xrightarrow{M_t,N_t,M_r,N_r\to\infty}\sqrt{M_{t}N_{t}M_{r}N_{r}}\Omega_{t}\Omega_{r}\mathbf{t}_u(f_u^w,d_u),
\end{align}
where $$\mathbf{t}_u(f_u^w,d_u)=[\mathbf{0}_{1\times\sum_{u'=0}^{u-1}K_{u'}},\bm{\eta}_{u,L}(f_u^w,d_u),\mathbf{0}_{1\times\sum_{u'=u+1}^{U^w}K_{u'}}],$$ with $\bm{\eta}_{u,L}(f_u^w,d_u) = [\eta_{u,1,L}(f_u^w,d_u),\ldots,\eta_{u,K_{q_u},L}(f_u^w,d_u)]\in\mathbb{C}^{1\times K_{u}}$ denoting the complex gain of the LOS path between user $u$ and subarray $k\in\mathcal{K}_{q_u}$ of the AP in the subcarrier $w$. $K_u$ represents the number of antenna subarrays allocated to user $u$. Note that $|\eta_{u,k,L}(f_u^w,d_u)|,k=1,\ldots,K_{u}$ are the same and denoted as $|\eta_{u,L}(f_u^w,d_u)|$. Obviously, the equivalent channel $\tilde{\mathbf{h}}_{u}(f_{u}^{w},d_{u})$ approaches some determined vector as the number of subarray antennas goes to infinity.

With the above characterization, we have the following result on the lower-bound of the achievable rate.
\begin{Lem}\label{theo:infrate}
When the number of subarray antennas approaches infinity, the lower-bound on the achievable rate of the indoor THz communication systems with one-bit DACs is given by
\begin{align}\label{eq:infrate}
R^o= \sum_{u=1}^{U}\sum_{w\in{\mathcal{W}_u}}B\log_{2}\left(1+\frac{K_{u}^2|\eta_{u,L}(f_u^w,d_u)|^{2}}{\xi_{u}^w}\right),
\end{align}
where
\begin{align}\label{eq:xi}
\xi_{u}^w = \mathbf{t}_u(f_u^w,d_u)\mathbf{C}_{0}{\mathbf{t}^H_u(f_u^w,d_u)},
\end{align}
with
\begin{align}\label{eq:cddinf}
\mathbf{C}_{0}& = \big[\arcsin(\text{diag}(\bm{\Xi}\bm{\Xi}^{H})^{-\frac{1}{2}} \Re\{\bm{\Xi}\bm{\Xi}^{H}\} \text{diag}(\bm{\Xi}\bm{\Xi}^{H})^{-\frac{1}{2}}) \nonumber\\
 &+j \arcsin(\text{diag}(\bm{\Xi}\bm{\Xi}^{H})^{-\frac{1}{2}}\Im\{\bm{\Xi}\bm{\Xi}^{H}\}\text{diag}(\bm{\Xi}\bm{\Xi}^{H})^{-\frac{1}{2}})\big]\nonumber\\
 &\hspace{2cm}-\Delta\bm{\Xi}\bm{\Xi}^H\Delta^H,
\end{align}
where
\begin{small}
\begin{align}
&\bm{\Xi}=\left[\begin{array}{llll}
\bm{\eta}_{1,L}^H(f_1^w,d_1) & \mathbf{0}_{ K_{1}\times1} & \cdots & \mathbf{0}_{K_{1}\times1} \\
\mathbf{0}_{ K_{2}\times1} & \bm{\eta}_{2,L}^H(f_2^w,d_2) &\cdots & \mathbf{0}_{K_{2}\times1} \\
\vdots & \vdots &\ddots &\vdots\\
\mathbf{0}_{ K_{U^w}\times1} & \mathbf{0}_{ K_{U^w}\times1} & \cdots & \bm{\eta}_{U^w,L}^H(f_{U^w}^w,d_{U^w})
\end{array}\right],\label{eq:ximat}\\
 &\bm{\Delta}= \text{diag}\bigg(\Big[\frac{1}{|\eta_{1,1,L}|},\ldots,\frac{1}{|\eta_{1,K_{1},L}|},\nonumber\\
 &\hspace{1cm}\ldots,\frac{1}{|\eta_{U^w,1,L}|},\ldots,\frac{1}{|\eta_{U^w,K_{U^w},L}|}\Big]\bigg).\label{eq:deltamat}
\end{align}
\end{small}
\end{Lem}
\emph{Proof:} See Appendix \ref{app:infrate} for details.\hfill$\square$

\begin{Rem}
As indicated in the proof, the inter-user interference can be eliminated by the analog transmit beamforming, where different sets of subarrays are allocated to different groups of users in the large subarray antenna regime and the data stream for each user can be forwarded solely to the allocated subarrays. That is, we have single-user transmission with the allocated subarrays of antennas. 
\end{Rem}

\begin{Rem}
Note also that (\ref{eq:infrate}) is a constant and is independent of the transmit power $P$. This tells us that the THz communication systems with one-bit DACs should work in the low power regime to be power efficient when the number of antennas for each subarray goes to infinity since increasing transmit power doesnot improve the performance. Moreover, (\ref{eq:infrate}) implies that the major error source is due to the one-bit DACs rather than the AWGN at the receiver.
\end{Rem}

\subsection{Single-User Case}
In the following, we consider the single-user case. Suppose that $P^w=\frac{P}{W}$ with $W$ representing the total number of subcarriers for user $u$. Note that the lower-bound on the achievable rate can be expressed similar to (\ref{eq:infrate}). Here, we consider a simple case and can derive the following result regarding the single-user case.
\begin{Corr}
In the single-user indoor THz communication systems, the lower-bound on the achievable rate for the user with $\psi_L^r=\phi_L^r=\psi_{L}^{t}=\phi_{L}^{t}=0$ in the large subarray antenna regime is given by
\begin{align}\label{eq:ratesu}
R^{o} =BW\log_{2}\left(1+\frac{1}{\frac{\pi}{2}-1}\right).
\end{align}
\end{Corr}
\emph{Proof: }Note that with the assumption on the AoD/AoA, we know that $\eta_{u,k,L}(f_u^w,d_u)=\eta_{u,L}(f_u^w,d_u),\forall k$, i.e., $\bm{\eta}_{u,L}(f_u^w,d_u)=\eta_{u,L}(f_u^w,d_u)\mathbf{1}_{1\times K}$. Now, $\mathbf{C}_{0}$ for one user takes the form similar to (\ref{eq:cddinf}) with the updated $\bm{\eta}_{u,L}(f_u^w,d_u)$, and can be expressed as
\begin{align}
\mathbf{C}_{0} = \left(\frac{\pi}{2}-1\right)\mathbf{1}_{K\times K}.
\end{align}
Therefore, we have
\begin{align}\label{eq:xis}
\xi_{u}^w &= \bm{\eta}_{u,L}(f_u^w,d_u)\mathbf{C}_{dd}\bm{\eta}^H_{u,L}(f_u^w,d_u),\nonumber\\
&= (\frac{\pi}{2}-1)|\eta_{u,L}(f_u^w,d_u)|^2K^2.
\end{align}
Considering the rate expression in (\ref{eq:infrate}) and (\ref{eq:xis}), we have the result in (\ref{eq:ratesu}).\hfill$\square$

\begin{Rem}
This tells us that the achievable rate for the user with zero AoD/AoA angles is insensitive to the distance between the AP and the number of subarrays in the large subarray antenna regime. That is, with the one-bit DACs, increasing the number of subarrays may not improve the spectral efficiency for certain user since the distortion introduced by the one-bit DACs increases with the number of subarrays as well.
\end{Rem}

In addition, there may be phase uncertainties for the THz band, which will degrade the achievable rate \cite{indoor}. Denote $\sigma_{m_{t},n_{t}}(f)$ and $\sigma_{m_{r},n_{r}}(f)$ as the random phase errors at the transmitter and receiver, respectively. Assume that $\sigma_{m_{t},n_{t}}(f)$ and $\sigma_{m_{r},n_{r}}(f)$ follow uniform distribution in $[-\varepsilon_{t}(f),\varepsilon_{t}(f)]$ and $[-\varepsilon_{r}(f),\varepsilon_{r}(f)]$, respectively. Then, we have the following proposition.
\begin{Lem1}\label{prop:phaseerror}
When the number of antennas for each subarray goes to infinity, the rate loss due to the phase uncertainties can be negligible in the indoor THz communications with one-bit DACs.
\end{Lem1}
\emph{Proof:} See Appendix \ref{app:phaseerror} for details.\hfill$\square$
\begin{Rem}
The result is different from \cite{indoor}, where infinite-resolution DACs are applied and the rate loss increases with the subarray size. This is generally because that the large distortion introduced by the one-bit DACs dominates the interference component.
\end{Rem}

\section{Numerical Results}

In this section, we numerically evaluate the performance of the indoor THz communication systems with one-bit DACs. The simulation parameters are given in Tables I and II. In addition, we assume the same size for each subarray, i.e., $M_t=N_t=M_r=N_r$.

\begin{table}
\caption{Transmission Windows}
\begin{center}
\begin{tabular}{|c|c|c|c|c|}

\hline
        & Distance &  \tabincell{c}{Transmission \\Windows (THz)} & User Group  \\
\hline
User1 & 10m & \tabincell{c}{0.6-0.7, \\0.8-0.95} & group1  \\
\hline
User2 & 5m & \tabincell{c}{0.6-0.725,\\ 0.8-0.925} & group2  \\
\hline
User3 & 1m & \tabincell{c}{0.5-0.6, \\0.7-0.8, 0.95-1} & group1  \\
\hline
\end{tabular}
\end{center}
\end{table}

 \begin{table}
 \caption{System Parameters}
 \begin{center}
 \begin{tabular}{ | l | l | }
 \hline Parameters& Values \\
 \hline
 $\Omega_{t}, \Omega_{r}$ & 20 dBi, 20 dBi \\
 \hline
 $B$ & 5 GHz\\
 \hline
 $K$ & 8\\
 \hline
$N_{0}$ & -75 dBm\\
  \hline
  $\varepsilon_{t}(f),\varepsilon_{r}(f)$ & $\frac{\pi}{18}, \frac{\pi}{18} $\\
   \hline
 \end{tabular}
\end{center}
\end{table}

\begin{figure}
    \centering
    \includegraphics[width=\figsize\textwidth]{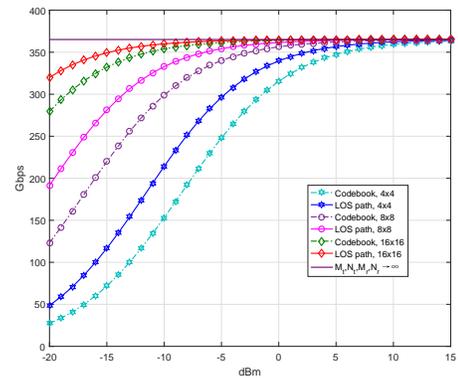}
    \caption{Comparison of different analog beamforming method.}
    \label{fig:losangle}
\end{figure}


Fig. \ref{fig:losangle} shows the achievable rate with different analog beamforming method as the transmit power increases for different subarray antenna size in the single-user case. We choose the analog beamforming angles either through codebook, i.e., (\ref{eq:equihbg}), or the direction of the LOS path, i.e., (\ref{eq:angles}). From the figure, as the transmit power increases, the curves with finite subarray size approach the constant value achieved with $M_t,N_t,M_r,N_r\to\infty$, which is independent of transmission power. Also, we can see that as the number of antennas for each subarray increases, the gap between the achievable rate with the analog beamforming specified by the LOS path and the one achieved by the codebook diminishes. In the following, we assume that the analog beamforming angles are specified by the AoD/AoA of the LOS path.

\begin{figure}
    \centering
    \includegraphics[width=\figsize\textwidth]{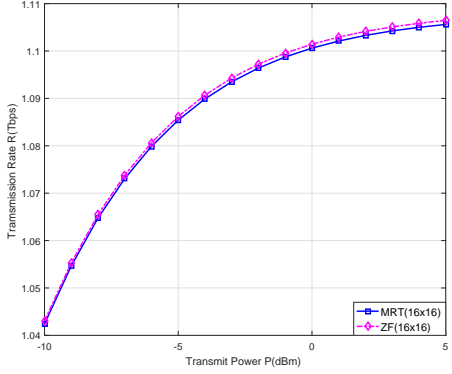}
    \caption{Comparison of the MRT and ZF.}
    \label{fig:mu}
\end{figure}

In Fig. \ref{fig:mu}, we plot the achievable rate with $M_t=N_t=M_r=N_r=16$ for the multi-user case. We assume that $K_1=5$ and $K_2=3$, i.e., 5 subarrays are allocated to group 1 while 3 subarrays are allocated to group 2. From the figure, the performance gap between ZF and MRT vanishes, which aligns with Theorem \ref{theo:infrate}.

\begin{figure}
   \centering
    \includegraphics[width=\figsize\textwidth]{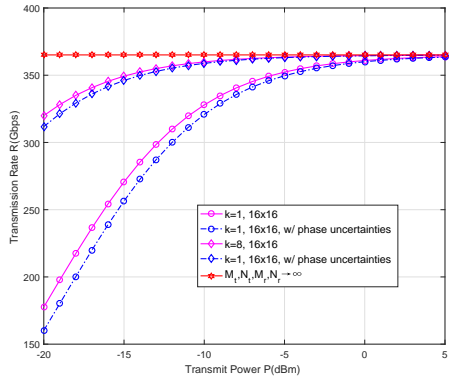}
     \caption{The single-user case with $\psi_L^r=\phi_L^r=\psi_{L}^{t}=\phi_{L}^{t}=0$.}
    \label{fig:phaseerror}
\end{figure}

In Fig. \ref{fig:phaseerror}, we plot the achievable rate with different numbers of antenna subarrays in the presence of phase uncertainties for the single-user with $\psi_L^r=\phi_L^r=\psi_{L}^{t}=\phi_{L}^{t}=0$. It can be seen from the figure that the different curves approach the same value achieved with $M_t,N_t,M_r,N_r\to\infty$, i.e., the impact of phase uncertainties on the achievable rate can be negligible in the large subarray antenna regime. In addition, we can see that $K=1$ achieves the same rate with $K=8$ in the limit as $P$ approaches infinity. Then, increasing the number of subarrays may not be helpful when the number of the antennas for each subarray is large enough.

 \section{CONCLUSIONS}

In this paper, we have investigated the indoor THz communications with one-bit DACs in the large subarray antenna regime. We have assumed that distance-aware multi-carrier strategy and antenna-of-subarrays architecture with hybrid precoding scheme are employed. We have proved that the analog beamforming angles for each group of users are decided the LOS path in the large subarray antenna regime. We have derived the closed-form expression for the lower bound on the achievable rate, which does not change with the transmit power. We have shown that the achievable rate for the single-user transmissions with one-bit DACs is robust to the phase uncertainties in the large subarray antenna regime. We have also provided the numerical results justifying the analysis.

\appendix
\subsection{Proof of Proposition \ref{prop:angles}}\label{app:angles}
Substituting (\ref{eq:channel}) into (\ref{eq:equivchannel1}), the effective channel between the subarray $k\in\mathcal{K}$ at the transmitter with analog beamforming angle $(\zeta_{t},\theta_{t})$ and a subarray at the user with analog beamforming angle $(\zeta_{r},\theta_{r})$ can be expressed as
 \begin{align}\label{eq:channelvalue}
 \hat{h}_k&=\mathbf{a}_{r}^{H}(\zeta_{r},\theta_{r})\mathbf{F}_{sub}(f,d)\mathbf{a}_{t}(\zeta_{t},\theta_{t})\nonumber\\
   &=\sqrt{M_{t}N_{t}M_{r}N_{r}}(\eta_{L}(f,d)\Omega_{t}\Omega_{r}\mathbf{a}_{r}^{H}(\zeta_{r},\theta_{r})\mathbf{a}_{r}(\psi_{L}^{r},\phi_{L}^{r})\nonumber\\
   &\times\mathbf{a}_{t}^{H}(\psi_{L}^{t},\phi_{L}^{t})\mathbf{a}_{t}(\zeta_{t},\theta_{t})+\sum_{i=1}^{J}\eta_{i}(f,d)\Omega_{t}\Omega_{r}\nonumber\\
 &\times\mathbf{a}_{r}^{H}(\zeta_{r},\theta_{r})\mathbf{a}_{r}(\psi_{i}^{r},\phi_{i}^{r})\mathbf{a}_{t}^{H}(\psi_{i}^{t},\phi_{i}^{t}))\mathbf{a}_{t}(\zeta_{t},\theta_{t}).
   \end{align}
Note that when $M_t, N_t\to\infty$, we have
\begin{small}
\begin{align}
&\lim_{M_{t}, N_{t}\rightarrow \infty} \left|\mathbf{a}^{H}_{t}(\psi_{L}^{t},\phi_{L}^{t})\mathbf{a}_{t}(\zeta_{t},\theta_{t})\right| \nonumber\\
&  =\lim_{M_{t}, N_{t}\rightarrow \infty}\bigg|\frac{1}{M_{t}N_{t}} e^{-j\frac{M_{t}-1}{2}[\frac{2\pi a}{\lambda}(\beta_{L1}(\zeta_{t},\theta_{t}))]} \nonumber\\
  &\times\frac{\sin[\frac{M_{t}\pi a}{\lambda}(\beta_{L1}(\zeta_{t},\theta_{t}))]}{\sin\left(\frac{\pi a}{\lambda}(\beta_{L1}(\zeta_{t},\theta_{t}))\right)}e^{-j\frac{N_{t}-1}{2}[\frac{2\pi a}{\lambda}(\beta_{L2}(\zeta_{t},\theta_{t}))]} \nonumber\\
 &\times \frac{\sin[\frac{N_{t}\pi a}{\lambda}(\beta_{L2}(\zeta_{t},\theta_{t}))]}{\sin\left(\frac{\pi a}{\lambda}(\beta_{L2}(\zeta_{t},\theta_{t}))\right)}\bigg|\\
  &=\delta(\beta_{L1}(\zeta_{t},\theta_{t}))\delta(\beta_{L2}(\zeta_{t},\theta_{t})),\label{eq:angleinf0}
  \end{align}
  \end{small}
 where
 \begin{align}
 \beta_{L1}(\zeta_{t},\theta_{t})=\cos\zeta_{t}\sin\theta_{t}-\cos\psi_{L}^{t}\sin\phi_{L}^{t},\label{eq:beta1}\\
 \beta_{L2}(\zeta_{t},\theta_{t})=\sin\zeta_{t}\sin\theta_{t}-\sin\psi_{L}^{t}\sin\phi_{L}^{t}.\label{eq:beta2}
 \end{align}
Similarly, we can show that
\begin{small}
\begin{align}
  &\hspace{-.7cm}\lim_{M_{r}, N_{r}\rightarrow \infty}\left|\mathbf{a}_{r}^{H}(\zeta_{t},\theta_{t})\mathbf{a}_{r}(\psi_{L}^{r},\phi_{L}^{r})\right| =\delta(\beta_{L1}(\zeta_{r},\theta_{r}))\delta(\beta_{L2}(\zeta_{r},\theta_{r})),\label{eq:angleinf1}\\
  &\hspace{-.6cm}\lim_{M_{t}, N_{t}\rightarrow \infty}\left|\mathbf{a}^{H}_{t}(\psi_{i}^{t},\phi_{i}^{t})\mathbf{a}_{t}(\zeta_{t},\theta_{t})\right|=\delta(\beta_{i1}(\zeta_{t},\theta_{t}))\delta(\beta_{i2}(\zeta_{t},\theta_{t})),\\
 &\hspace{-.6cm}\lim_{M_{r}, N_{r}\rightarrow \infty}\left|\mathbf{a}_{r}^{H}(\zeta_{t},\theta_{t})\mathbf{a}_{r}(\psi_{i}^{r},\phi_{i}^{r})\right| =\delta(\beta_{i1}(\zeta_{r},\theta_{r}))\delta(\beta_{i2}(\zeta_{r},\theta_{r})),\label{eq:angleinf3}
\end{align}
\end{small}
where
 \begin{align}
 \beta_{L1}(\zeta_{r},\theta_{r})&=\cos\zeta_{r}\sin\theta_{r}-\cos\psi_{L}^{r}\sin\phi_{L}^{r},\\
 \beta_{L2}(\zeta_{r},\theta_{r})&=\sin\zeta_{r}\sin\theta_{r}-\sin\psi_{L}^{r}\sin\phi_{L}^{r},
  \end{align}
   \begin{align}
    \beta_{i1}(\zeta_{t},\theta_{t})&=\cos\zeta_{t}\sin\theta_{t}-\cos\psi_{i}^{t}\sin\phi_{i}^{t},\\
  \beta_{i2}(\zeta_{t},\theta_{t})&=\sin\zeta_{t}\sin\theta_{t}-\sin\psi_{i}^{t}\sin\phi_{i}^{t},\\%
 \beta_{i}(\zeta_{r},\theta_{r})&=\cos\zeta_{r}\sin\theta_{r}-\cos\psi_{i}^{r}\sin\phi_{i}^{r},\\
 \beta_{i2}(\zeta_{r},\theta_{r})&=\sin\zeta_{r}\sin\theta_{r}-\sin\psi_{i}^{r}\sin\phi_{i}^{r}.
 \end{align}
Taking the limit of (\ref{eq:channelvalue}) as $M_t,N_t,M_r,N_r$ approach infinity and substituting (\ref{eq:angleinf0}), (\ref{eq:angleinf1})-(\ref{eq:angleinf3}) into the resulting equation, it can be easily verified that (\ref{eq:equihbg}) is maximized with the choice of $(\zeta_{t},\theta_{t})=(\psi_{L}^{t},\phi_{L}^{t})$ and $(\zeta_{r},\theta_{r}) = (\psi_{L}^{r},\phi_{L}^{r})$ since $\parallel \mathbf{h}\parallel^{2}=\sum_{k\in\mathcal{K}}|\hat{h}_k|^2$ and $|\eta_L(f,d)|>|\eta_i(f,d)|,\forall\,i$.\hfill$\square$

\subsection{Proof of Theorem \ref{theo:infrate}}\label{app:infrate}

 First, we can see from (\ref{eq:hmat}), (\ref{eq:infchannel}), (\ref{eq:mrt}) and (\ref{eq:zf}) that
\begin{align}\label{eq:infQ}
&\mathbf{Q}_{MRT}=\mathbf{Q}_{ZF}=\sqrt{\frac{P^w}{\sum_{u\in\mathcal{U}^w}\parallel\bm{\eta}_{u,L}\parallel^2}}\bm{\Xi},
\end{align}
where $\bm{\Xi}$ is defined in (\ref{eq:ximat}).

Substituting (\ref{eq:infQ}) into (\ref{eq:quantmat}) gives us
\begin{align}\label{eq:infG}
 \mathbf{G}&=\sqrt{\frac{2}{\pi K^w}\sum_{u\in\mathcal{U}^w}\parallel\bm{\eta}_{u,L}\parallel^2} \bm{\Delta},
 \end{align}
 where $\bm{\Delta}$ is defined in (\ref{eq:deltamat}).

Combining (\ref{eq:infQ}) and (\ref{eq:infG}) with (\ref{eq:rate}), and defining $\xi_{u}^w$ in (\ref{eq:xi}), we can arrive at (\ref{eq:infrate}) by noting that $\frac{N_0}{M_tN_tM_rN_r}\to0$.\hfill$\square$

\subsection{Proof of Proposition \ref{prop:phaseerror}}\label{app:phaseerror}
The idea of this proof is similar to \cite[Section V]{indoor}. We include the proof here for the sake of the reader. We can first obtain the radiation pattern of one antenna subarray associated with the transmit beamforming vector specified by $(\tilde{\psi_q^{t}},\tilde{\phi}_q^{t})$ for the AP as
\begin{align}
&|A(\phi,\theta,f)|^{2}=A(\phi,\theta,f)A^{H}(\phi,\theta,f)\nonumber\\
&=\frac{1}{MN}\sum_{m=0}^{M-1}\sum_{n=0}^{N-1}\sum_{m'=0}^{M-1}\sum_{n'=0}^{N-1}e^{j(\sigma_{m,n}(f)-\sigma_{m',n'}(f))}\nonumber\\
&\times e^{j\frac{2\pi a}{\lambda}(m-m')\beta_1+(n-n')\beta_2},
\end{align}
where $\beta_1=\cos\tilde{\psi_q^{t}}\sin\tilde{\phi}_q^{t}-\cos\phi\sin\theta$ and $\beta_2=\sin\tilde{\psi_q^{t}}\sin\tilde{\phi}_q^{t}-\sin\phi\sin\theta$.
Since $\sigma_{m,n}(f)-\sigma_{m',n'}$ is very small, we can apply the Taylor's expansion and have
\begin{align}
e^{j(\sigma_{m,n}(f)-\sigma_{m',n'}(f))}&\approx 1+j(\sigma_{m,n}(f)-\sigma_{m',n'}(f))\nonumber\\
&\hspace{.5cm}-\frac{1}{2}(\sigma_{m,n}(f)-\sigma_{m',n'}(f))^{2}.
\end{align}
Then the average radiation pattern of one antenna subarray at the transmitter can be written as
\begin{small}
\begin{align}
\hspace{-.3cm}\mathbb{E}[|A_{t}(\phi,\theta,f)|^{2}]=|A_{t}(\phi,\theta)|^{2}-\frac{\varepsilon_{t}^{2}(f)}{3}|A_{t}(\phi,\theta)|^{2}+\frac{\varepsilon_{t}^{2}(f)}{3}.
\end{align}
\end{small}
Similarly, the average radiation pattern at the receiver is
\begin{small}
\begin{align}
\hspace{-.3cm}\mathbb{E}[|A_{r}(\phi,\theta,f)|^{2}]=|A_{r}(\phi,\theta)|^{2}-\frac{\varepsilon_{r}^{2}(f)}{3}|A_{r}(\phi,\theta)|^{2}+\frac{\varepsilon_{r}^{2}(f)}{3}.
\end{align}
\end{small}
Then, the equivalent channel in (\ref{eq:infchannel}) becomes
\begin{small}
\begin{align}\label{eq:infchannelerror}
&\tilde{\mathbf{h}}_{u}(f_{u}^{w},d_{u})\xrightarrow{M_t,N_t,M_r,N_r\to\infty}\nonumber\\
&\sqrt{M_{t}N_{t}-(M_tN_t-1)\frac{\varepsilon_{t}^{2}(f)}{3} + \frac{\varepsilon_{t}^{2}(f)}{3}}\nonumber\\
&\times\sqrt{M_{r}N_{r} - (M_rN_r-1)\frac{\varepsilon_{r}^{2}(f)}{3}+\frac{\varepsilon_{r}^{2}(f)}{3}}\Omega_{t}\Omega_{r}\mathbf{t}_u(f_u^w,d_u).
\end{align}
\end{small}
Incorporating (\ref{eq:infchannelerror}) with Appendix \ref{app:infrate} gives us the same value in (\ref{eq:infrate}) as $M_t,N_t,M_r,N_r\to\infty$, implying zero rate loss.\hfill$\square$

\end{document}